\documentclass[11pt]{article}

\usepackage{graphics}

\newcommand{\ket}[1]{\vert#1\rangle}

\begin{document}

\title{Circuit for Shor's algorithm using 2n+3 qubits}

\author{St\'{e}phane Beauregard\thanks{D\'{e}partement de Physique et D\'{e}partement d'Informatique et de Recherche Op\'{e}rationnelle, Universit\'{e} de Montr\'{e}al, {\tt beaurest@iro.umontreal.ca}. Supported by NSERC.}}

\date{}

\maketitle

\begin{abstract}

We try to minimize the number of qubits needed to factor an integer of $n$ bits using Shor's algorithm on a quantum computer. We introduce a circuit which uses $2n+3$ qubits and $O(n^3 lg(n))$ elementary quantum gates in a depth of $O(n^3)$ to implement the factorization algorithm. The circuit is computable in polynomial time on a classical computer and is completely general as it does not rely on any property of the number to be factored.

\end{abstract}

\section{Introduction}
\noindent
Since Shor discovered a polynomial time algorithm for factorization on a quantum computer~\cite{Shor}, a lot of effort has been directed towards building a working quantum computer. Despite all these efforts, it is still extremely difficult to control even a few qubits. It is thus of great interest to study exactly how few qubits are needed to factor an $n$-bit number.

\begin{figure}[htbp]
\begin{center}
\includegraphics{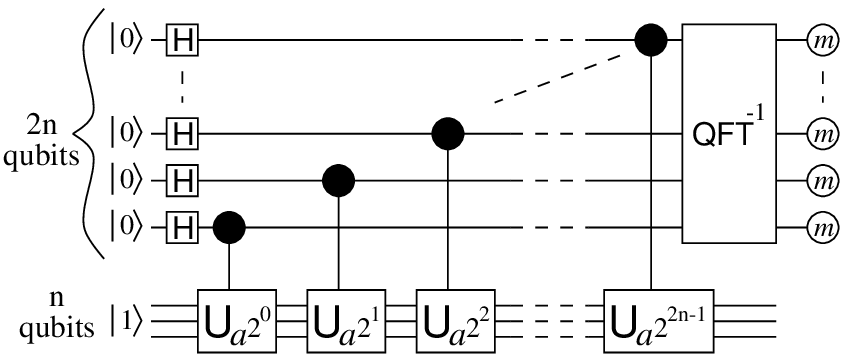}
\caption{\small{The order-finding circuit for quantum factorization. $\textbf{U}_a$ implements $\ket{x} \to \ket{(ax)\textrm{mod~}N}$ and the measurements followed by classical postprocessing yields the order $r$ of $a$ modulo $N$ with good probability.}}\label{circuit}
\end{center}
\end{figure}

Quantum factorization consists of classical preprocessing, a quantum algorithm for order-finding and classical postprocessing~\cite{Shor, Cleve, CN} (fig.~\ref{circuit}). We will concentrate on the quantum part of factorization and consider classical parts as being free as long as they are computable in polynomial time. The only use of quantum computation in Shor's algorithm is to find the order of $a$ modulo $N$, where $N$ is an $n$-bit integer that we want to factor. The order $r$ of $a$ modulo $N$ is the least positive integer such that $a^r \equiv 1($mod $N)$.

\setcounter{footnote}{0}
\renewcommand{\thefootnote}{\alph{footnote}}

For completeness, we now give the full algorithm for factoring $N$ as given in~\cite{CN}:
\begin{enumerate}
\item If $N$ is even, return the factor 2.
\item Classically determine if $N=p^q$ for $p \geq 1$ and $q \geq 2$ and if so return the factor $p$ (this can be done in polynomial time).
\item Choose a random number $a$ such that $1 < a \leq N-1$. Using Euclid's algorithm, determine if gcd($a,N$)$> 1$ and if so, return the factor gcd($a,N$).
\item Use the order-finding quantum algorithm to find the order $r$ of $a$ modulo $N$.
\item If $r$ is odd or $r$ is even but $a^{r/2}= -1($mod $N)$, then go to step (iii). Otherwise, compute gcd($a^{r/2}-1,N$) and gcd($a^{r/2}+1,N$). Test to see if one of these is a non-trivial factor of $N$, and return the factor if so. 
\end{enumerate}

It can be shown that with probability at least one half, $r$ will be even and $a^{r/2} \neq -1($mod $N)$~\cite{Shor, CN}. The quantum part of the algorithm (step 4) is known to be computable in polynomial time on a quantum computer. Using classical techniques, it is straigthforward to build the order-finding circuit (fig.~\ref{circuit}) using a polynomial number of elementary gates and a linear number of qubits~\cite{Shor}. Because the depth of the circuit is related to its running time, it is desirable to minimize this depth, and much progress has been made in that direction~\cite{CW}. We propose to take the problem from the other side: by how much can the number of qubits be reduced for factorization in polynomial time? Answering this question would give insights on the size of a quantum computer useful for factorization. We thus introduce a new order-finding circuit focused on reducing the number of qubits while still using only a polynomial number of elementary quantum gates. We also somewhat try to minimize the depth of the circuit, but very little parallelization is available since we avoid using any unnecessary qubit. 

\begin{figure}[htbp]
\begin{center}
\includegraphics{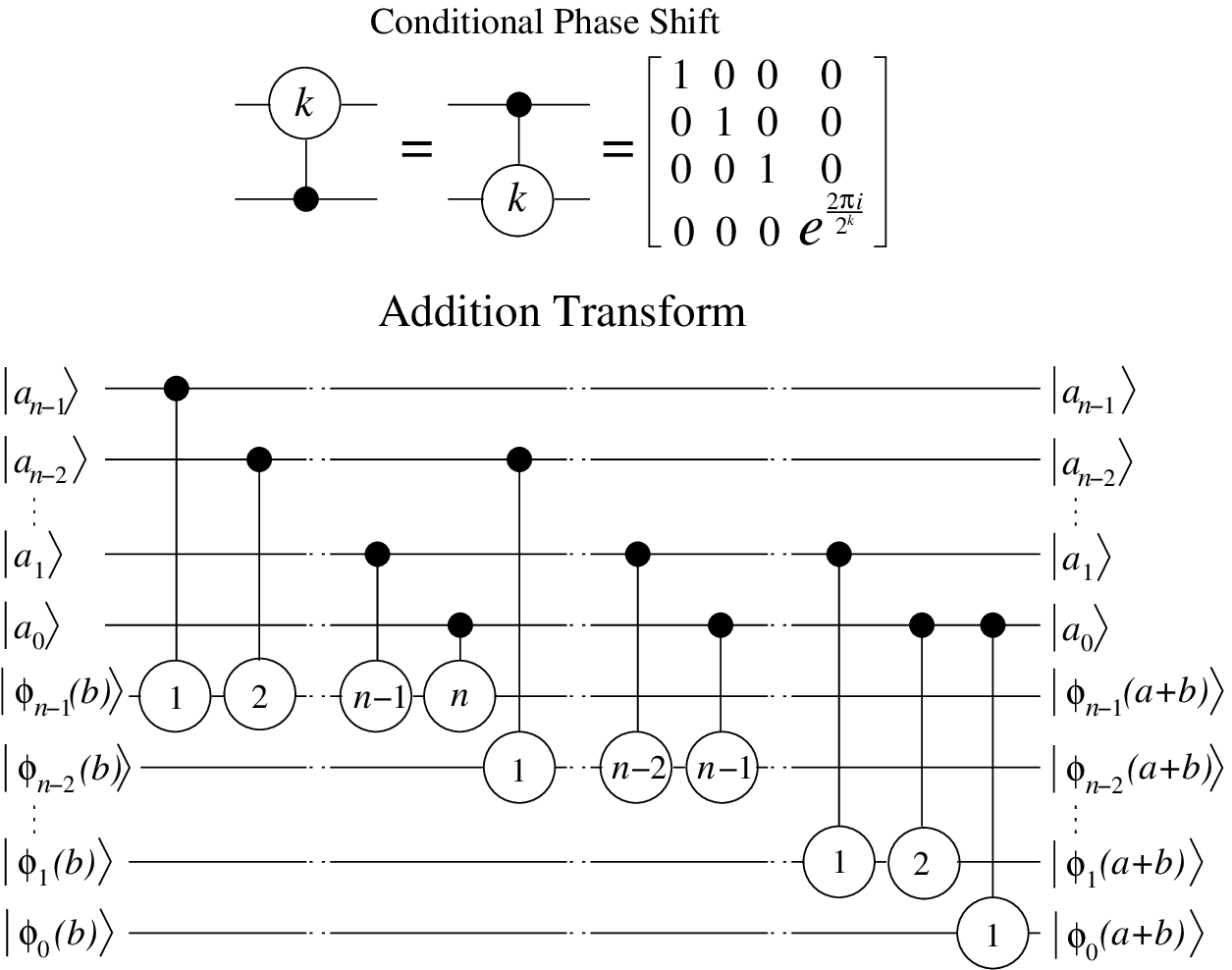}
\caption{\small{The quantum addition as described by Draper~\cite{Draper}.}}\label{draperadd}
\end{center}
\end{figure}

\section{The Circuit}
\noindent
The circuit for factorization that will be discussed here was inspired in part by a circuit from Vedral, Barenco and Ekert~\cite{VBE}. To reduce the number of qubits, we use a variant of a quantum addition algorithm described by Draper~\cite{Draper} (fig.~\ref{draperadd}). Other techniques used to reduce the number of qubits are the hardwiring of classical values and the sequential computation of the Fourier transform.

The quantum addition of figure~\ref{draperadd} takes as input $n$ qubits representing a number $a$, and $n$ more qubits containing the quantum Fourier transform of an other number $b$, denoted by $\phi(b)$. After the addition, the first register keeps the same value $a$ but the bottom register now contains the quantum Fourier transform of $(a+b)$mod~$2^n$, denoted by $\phi(a+b)$.

\begin{figure}[htbp]
\begin{center}
\includegraphics{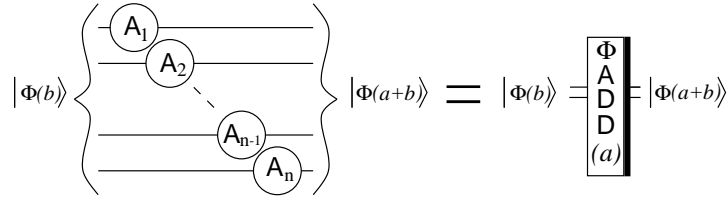}
\caption{\small{The circuit for addition of a classical value $a$ to the quantum value $b$ in the Fourier space. The gates $\textbf{A}_i$ are classically computed combinations of phase shifts.}}
\label{phiadd}
\end{center}
\end{figure}

\subsection{The adder gate}
\noindent
Adding together two quantum registers is, however, more than we ask for. We are trying to find the period of the function $(a^x)$mod~$N$ where $a$ is a classical random number smaller than $N$. Since $a$ is classical, we only need to be able to add a classical value to a quantum register. We can thus change the qubits representing $a$ in figure~\ref{draperadd} to classical bits. The controlled gates are then classically controlled, and since we know what $a$ is beforehand, we might as well precompute the product of all gates on each single qubit and apply only one gate for every single qubit. These are one-qubit gates, which also makes them easier to implement.

Since the addition takes place in the Fourier space, we will call this circuit the $\phi ADD(a)$ gate where $a$ is the classical value added to the quantum register (fig.~\ref{phiadd}). Notice the thick black bar on the right, used to distinguish the gate from its unitary inverse. In order to prevent overflow, we need $n+1$ qubits for the quantum register instead of $n$, so that $\phi(b)$ is effectively the QFT of an ($n+1$)-qubit register containing a $n$-bit number (thus the most significant qubit before the QFT preceding the addition is always  $\ket{0}$).

\begin{figure}[htbp]

\begin{center}
\includegraphics{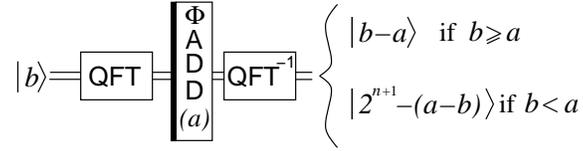}
\caption{\small{The effect of the reverse $\phi ADD(a)$ gate on $\ket{\phi(b)}$.}}\label{phiminus}
\end{center}
\end{figure}

If we apply the unitary inverse of the $\phi ADD(a)$ gate with input $\phi(b)$, we get either $\phi(b-a)$ if $b \geq a$, or $\phi(2^{n+1}-(a-b))$ if $b < a$. Thus if $b < a$, the most significant qubit of the result is always $\ket{1}$, whereas it is always $\ket{0}$ if $b \leq a$. This reverse $\phi ADD(a)$ gate can be useful for subtraction and comparison purposes (fig.~\ref{phiminus}) and we use a black bar on the left to distinguish it from the regular gate. The unitary inverse of a circuit is obtained by applying the unitary inverse of each elementary gate in reverse order.

\subsection{The modular adder gate}
\begin{figure}[htbp]
\begin{center}
\includegraphics{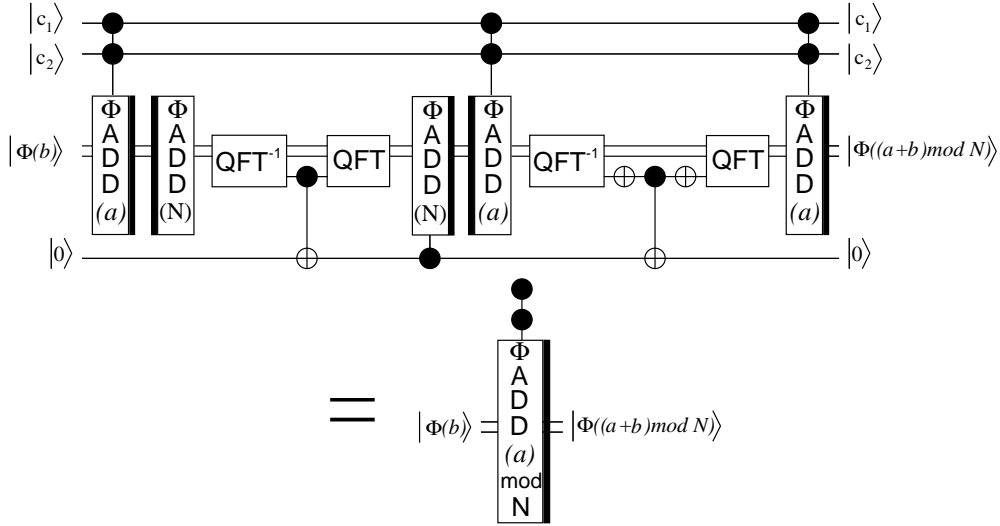}
\caption{\small{The doubly controlled $\phi ADD(a)MOD(N)$ gate with $c_1=c_2=1$. If either of the control qubits is in state $\ket{0}$, the output of the gate is $\ket{\phi(b)}$ since $b<N$.}}\label{ccaddmod}
\end{center}
\end{figure}

\noindent
Now that we have a $\phi ADD(a)$ gate, we can use it to build a modular adder gate (fig.~\ref{ccaddmod}). For future use, two control qubits are included in the circuit. For the modular adder gate, we need to compute $a + b$ and subtract $N$ if $a+b \geq N$. However, it is not so easy to implement this operation in a reversible way. The input to the $\phi ADD(a)MOD(N)$ gate is $\phi(b)$ with $b<N$, and the classical number $a$ that we add is also smaller than $N$.

We begin by applying a $\phi ADD(a)$ gate to the register $\phi(b)$. The quantum register now contains $\phi(a+b)$ with no overflow because we were careful enough to put an extra qubit in state $\ket{0}$ along with the value $b$ before applying the QFT. We next run a reverse $\phi ADD(N)$ to get $\phi(a+b-N)$. If $a+b < N$, we did not have to subtract $N$ but now we can determine if $a+b < N$ by checking the most significant bit of $a+b-N$. However, to access this most significant bit we need to inverse the QFT on the whole register containing $\phi(a+b-N)$. We can then use this qubit as the controlling qubit of a controlled-not gate acting on an ancillary qubit. It is then possible to reapply the QFT and use this ancilla as a control qubit for a $\phi ADD(N)$ controlled gate, so that if $a+b < N$ we add back the value $N$ that we subtracted earlier. We now have $\phi((a+b)$mod $N)$ in the register, and we are done except for the ancilla which is now a junk bit. We have to restore it to $\ket{0}$ somehow, otherwise the computation will not be clean and the algorithm will not work\footnote{Indeed, for the order-finding algorithm to work, we need to find the period of $(a^x)$mod~$N$ but the period of the garbage bits can be something else.}.

Restoring the ancilla to $\ket{0}$ is no easy task if we do not want to waste qubits. We can still do it by using the identity:

\begin{equation}
(a+b)\textrm{mod }N\geq a \Leftrightarrow a+b<N.
\end{equation}

Hence, we only have to compare $(a+b) $mod~$N$ with the value $a$ using essentially the same trick as before. We run an inverse $\phi ADD(a)$ followed by an inverse QFT to get the most significant qubit of $(a+b)$mod $N -a$. This qubit is $\ket{0}$ if $(a+b)$mod $N \geq a$. We apply a NOT gate on this qubit and use it as the controlling qubit of a controlled-not gate targeting the ancilla. The ancilla is thus restored to $\ket{0}$ and we can apply a NOT gate again on the control wire, followed by a QFT and a $\phi ADD(a)$ gate on the quantum register. After this, we have a clean computation of $(a+b)$mod~$N$ in the Fourier space. 

Again, what we need exactly is a doubly controlled version of the $\phi ADD(a)MOD(N)$ gate. In order to reduce the complexity of the circuit, we will doubly control only the $\phi ADD(a)$ gates instead of all the gates (fig.~\ref{ccaddmod}). If the $\phi ADD(a)$ gates are not performed, it is easy to verify that the rest of the circuit implements the identity on all qubits because $b<N$.

\begin{figure}[htbp]
\begin{center}
\includegraphics{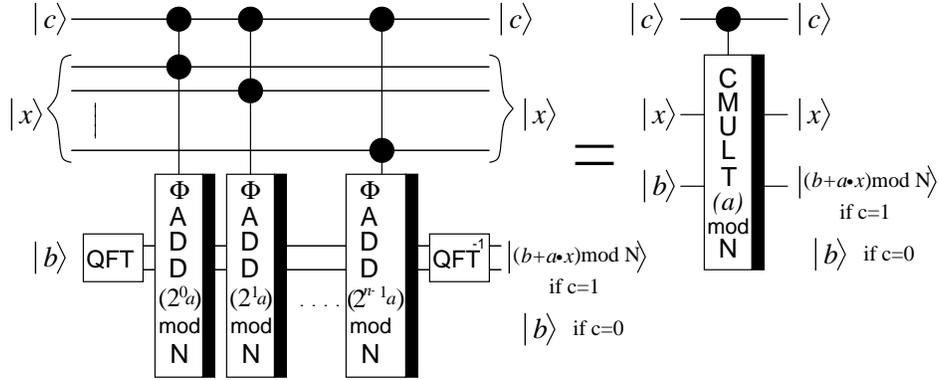}
\caption{\small{The $CMULT(a)MOD(N)$ gate.}}\label{cmul}
\end{center}
\end{figure}

\subsection{The controlled multiplier gate}
\noindent
The next step is to use the doubly controlled $\phi ADD(a)MOD(N)$ gate to build a controlled multiplier gate that we will call $CMULT(a)MOD(N)$ (fig.~\ref{cmul}). This gate takes three inputs, $\ket{c}\ket{x}\ket{b}$, and its output depends on the qubit $\ket{c}$. If $\ket{c} = \ket{1}$, the output is $\ket{c}\ket{x}\ket{b+(ax)\textrm{mod~}N}$. If $\ket{c}=\ket{0}$, then the input is unchanged and stays $\ket{c}\ket{x}\ket{b}$. This gate is very straightforward to implement using doubly controlled $\phi ADD(a)MOD(N)$ gates. We use the identity:

\begin{eqnarray}
& (ax)\textrm{mod~}N = & \nonumber \\ & (...((2^0ax_0)\textrm{mod~}N + 2^1ax_1)\textrm{mod~}N+... +2^{n-1}ax_{n-1})\textrm{mod~}N. &
\end{eqnarray}

\begin{figure}[htbp]
\begin{center}
\includegraphics{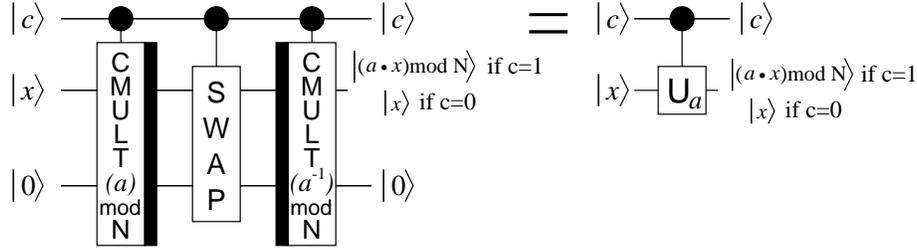}
\caption{\small{The controlled-$U_a$ gate.}}\label{ctrlu}
\end{center}
\end{figure}

Thus we only need $n$ successive doubly controlled modular adder gates, each of them adding a different value $(2^{i}a)\textrm{mod~}N$ with $0 \leq i < n$ to get the $CMULT(a)MOD(N)$ gate. We now have a controlled gate that takes $\ket{x}\ket{b}$ to $\ket{x}\ket{b+(ax)\textrm{mod~}N}$. What we would need instead is a controlled gate that takes $\ket{x}$ to $\ket{(ax)\textrm{mod~}N}$. This can however be obtained by a clever trick from reversible computing that uses two controlled multiplication gates (fig~\ref{ctrlu}). 

We first apply the $CMULT(a)MOD(N)$ gate to $\ket{c}\ket{x}\ket{0}$. We follow with a SWAP between the two registers if the qubit $\ket{c} = \ket{1}$ (that is effectively a controlled-SWAP on the registers)\footnote{We can do without the SWAP by modifying all later gates accordingly, but the SWAP simplifies the layout of the circuit without affecting the order of the complexity.}. We only need to control-SWAP $n$ qubits, not $n+1$. Indeed, the most significant qubit of $(ax)\textrm{mod~}N$ will always be 0 since we were careful to include one extra qubit to store the overflow in the $\phi ADD(a)$ gate. We then finish with the inverse of a $CMULT(a^{-1})MOD(N)$ circuit. The value $a^{-1}$, which is the inverse of $a$ modulo $N$, is computable classically in polynomial time using Euclid's algorithm and it always exists since gcd$(a,N)=1$. The fact that we apply the inverse of the circuit means that the circuit effectively takes $\ket{c}\ket{x}\ket{b}$ to $\ket{c}\ket{x}\ket{(b-a^{-1}x)\textrm{mod~}N}$.

The resulting gate will be called $C$-$U_a$ for controlled-$U_a$. It does nothing if $\ket{c}=\ket{0}$ but if $\ket{c}=\ket{1}$, then the two registers take the following values:

\begin{eqnarray}
\ket{x}\ket{0} \to \ket{x}\ket{(ax)\textrm{mod~}N} \to \ket{(ax)\textrm{mod~}N}\ket{x} \to \nonumber \\ \ket{(ax)\textrm{mod~}N}\ket{(x-a^{-1}ax)\textrm{mod~}N}=\ket{(ax)\textrm{mod~}N}\ket{0}.
\end{eqnarray}

Since the bottom register returns to $\ket{0}$ after the computation, we can consider this extra register as being part of the $C$-$U_a$ gate, thus the gate effectively takes $\ket{x}$ to $\ket{(ax)\textrm{mod~}N}$. This is exactly the gate we need to run the quantum order-finding circuit (fig~\ref{circuit}). Of course, we don't need to apply $C$-$U_a$ $n$ times to get $(C$-${U_a})^n$ because we can directly run $C$-$U_{a^n}$ (where $a^n\textrm{mod~}N$ in computed classically) which is the same as $(C$-${U_a})^n$ since:

\begin{eqnarray}
(a^nx)\textrm{mod~}N = \underbrace{(a...(a(ax)\textrm{mod~}N)\textrm{mod~}N...)\textrm{mod~}N}_{n\textrm{ times}}.
\end{eqnarray}

\subsection{The one controlling-qubit trick}
\noindent

\begin{figure}[htbp]

\begin{center}
\includegraphics{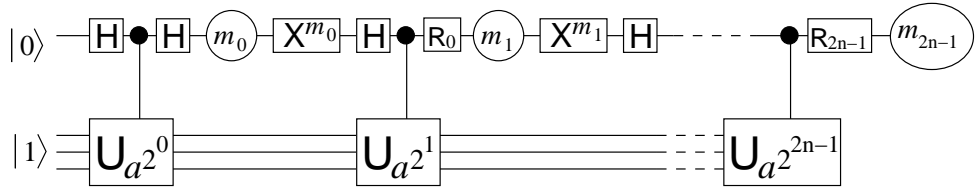}
\caption{\small{The one control qubit trick for factoring. The \textbf{R} gates depend on all previous measurement results and implement the inverse QFT, while the \textbf{X} gates are negations conditionned on the result of each measurement.}}\label{trick}
\end{center}
\end{figure}

An advantage of using the $C$-${U_{a^{2^j}}}$ gates for Shor's algorithm is the fact that we don't really need the total $2n$ controlling qubits. In fact, it can be shown that only one controlling qubit is sufficient~\cite{ME, Zalka, PP}. This is possible because the controlled-$U$ gates all commute and the inverse QFT can be applied semi-classically. Indeed, we can get all the bits of the answer sequentially as in figure~\ref{trick}. Each measured bit dictates which unitary transformation we have to apply after every controlled-$U$ step before the next measurement. This simulates the inverse QFT followed by a measurement on all qubits as in figure~\ref{circuit}. We save an important number of qubits this way, and in fact we need only a total of $2n+3$ qubits to factor an $n$-bit number as we will show in the complexity analysis section.

\begin{figure}[htbp]
\begin{center}
\includegraphics{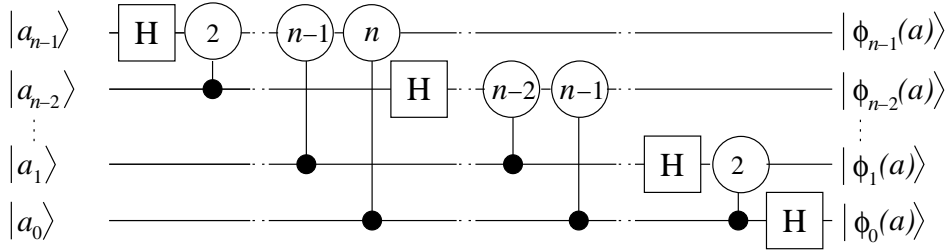}
\caption{The exact quantum Fourier transform. \textbf{H} is the Hadamard gate.}\label{qft}
\end{center}
\end{figure}

\subsection{The quantum Fourier transform}
\noindent
The implementation of the exact $QFT$ on $n$ qubits requires $O(n^2)$ operations~\cite{CN} (fig.~\ref{qft}). However, in physical implementations, there will always be a threshold for the precision of the gates. Since many phase shifts will be almost negligible, we will in practice ignore the ones with $k$ greater than a certain threshold $k_{max}$. This approximate $QFT$ is in fact very close to the exact $QFT$ even with $k_{max}$ logarithmic in $n$. In fact, it has been shown~\cite{Coppersmith} that the error introduced by ignoring all gates with $k>k_{max}$ is proportional to $n 2^{-k_{max}}$.We can thus choose $k_{max} \in O(\lg(\frac{n}{\epsilon}))$.

The implementation of the approximate $QFT$ on $n$ qubits requires $O(n\lg(n))$ gates. There seems to be no obvious way to reduce the depth of either the exact $QFT$ and the approximate $QFT$ on $n$ qubits below $O(n)$ without using extra qubits~\cite{Moore}. The depth of the QFT on $n+1$ qubits is thus $O(n)$ with the little parallelization available without extra qubits.

\subsection{The controlled-SWAP}
\noindent

\begin{figure}[htbp]
\begin{center}
\includegraphics{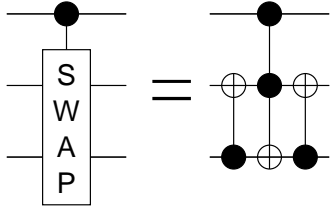}
\caption{The controlled-SWAP gate.}\label{cswap}
\end{center}
\end{figure}

The controlled-SWAP on one qubit is very easy to implement (fig.~\ref{cswap}). Only two controlled-not and one Toffoli are needed to perform the SWAP on two qubits controlled by a third. Thus, $O(n)$ gates are needed to control-SWAP $n$ qubits, that is, swap $n$ qubits with $n$ others with one control qubit.

\section{Complexity Analysis}
\noindent
We now analyze the complexity of the given circuit for performing factorization of an $n$-bit number $N$. The analysis keeps track of the number of qubits, the order of the number of gates and the order of the depth of the circuit. For the depth of the circuit, we consider that it will be possible to apply simultaneously different quantum gates that act on different qubits of the quantum computer. However, we consider impossible to have one qubit controlling many operations in the same step. The circuit uses only single qubit gates, up to doubly controlled conditionnal phase shifts and up to doubly controlled not gates. These gates can be implemented using a constant number of single qubit gates and controlled-nots~\cite{Barenco}, so they can all be considered as elementary quantum gates.

The $\phi ADD(a)$ circuit (fig.~\ref{phiadd}), where $a$ is a classical value, requires $n+1$ qubits and $O(n)$ single qubit gates in constant depth. The number of qubits is $n+1$ because we need an extra qubit to prevent overflows. When a control qubit is added to the circuit, the depth becomes $O(n)$ since the conditional phase shifts have to be done sequentially. Indeed, the control qubit has to control each phase shift one at a time. The doubly controlled $\phi ADD(a)MOD(N)$ circuit (fig.~\ref{ccaddmod}) requires $n+4$ qubits. It also requires $O(n k_{max})$ gates, but has a depth of only $O(n)$ regardless of $k_{max}$ because the QFTs can be somewhat parallelized. The $CMULT(a)MOD(N)$ circuit is only $n$ doubly controlled $\phi ADD(a)MOD(N)$ circuits. It thus takes $2n+3$ qubits, $O(n^2 k_{max})$ gates and a depth of $O(n^2)$ to implement the $CMULT(a)MOD(N)$ circuit. Two of these circuits along with the controlled-SWAP are needed for the $C$-$U_a$ circuit.  The controlled-SWAP on $n$ qubits requires only $O(n)$ gates and depth, so the $C$-$U_a$ circuit requires $2n+3$ qubits, $O(n^2 k_{max})$ gates and a depth of $O(n^2)$ again.

For the whole order-finding circuit, that is, the whole quantum part of Shor's algorithm, we need $2n$ of these $C$-$U_a$ circuits. The quantum resources needed are thus $2n+3$ qubits, $O(n^3 k_{max})$ gates and a depth of $O(n^3)$. If we decide to use the exact $QFT$ in the additions, then we would have $k_{max} = n$. As we argued earlier, this would not be clever because the implementation is sure to have hardware errors anyway. We thus should use the approximate $QFT$ with $k_{max} = O(\lg(\frac{n}{\epsilon})$), so that the number of gates is in $O(n^3 \lg(n))$ for any $\epsilon$ polynomial in $\frac{1}{n}$. 

This result of $2n+3$ qubits is slightly better than previous circuits for factorization. Vedral, Barenco and Ekert published a circuit of $7n+1$ qubits and $O(n^3)$ elementary gates for modular exponentiation~\cite{VBE}. It is mentionned that this number can be easily reduced to $5n+2$ qubits with basic optimization and further reduced to $4n+3$ if unbounded Toffoli gates ($n$-controlled nots) are available. Beckman, Chan, Devabhaktoni and Preskill provided an extended analysis~\cite{Beckman} of modular exponentiation, with a circuit of $5n+1$ qubits using elementary gates and $4n+1$ if unbounded Toffoli gates are available. Zalka also described a method for factorization with $3n+O(\lg{n})$ qubits using only elementary gates~\cite{Zalka}.

The availability of unbounded Toffoli gates will of course depend on the physical implementation of the quantum computer, but it is assumed throughout our design and analysis that such gates cannot be considered elementary. For that matter, if we do not restrict the type and size of the quantum gates in any way, order-finding can be achieved with $n+1$ qubits by directly using controlled multiplication gates~\cite{PP}. 

Of the $2n+3$ qubits used in the circuit provided here, one is used as an ancilla for modular addition, one is used to prevent addition overflows and $n$ are used as an ancillary register to get modular multiplication from successive additions. An order-finding circuit using elementary gates and less than $2n + O(1)$ qubits is not ruled out yet, but it seems that a different method would have to be used for modular multiplication to get such a circuit.

Fifteen is the smallest number on which Shor's algorithm can be applied. The circuit for factorization of $N=15$ uses eleven qubits as given here. However, the classical computation performed to build it gives a lot of information on the order of the number $a$. Indeed, for any $1<a<15$, the order of $a$ is either two or four. Most of the multiplications in the circuit are simply the identity and can be removed, which amounts to many unused qubits. The number 15 was factored using NMR with seven qubits in an impressive display of quantum control by Vandersypen, Steffen, Breyta, Yannoni, Sherwood and Chuang~\cite{VSB+}.

The importance of reducing the number of qubits versus reducing the depth of a quantum computation is not clear as quantum computers of useful size are not yet available. We have to keep in mind that error correction will most probably have to be used on quantum computers, which will create an overhead in the number of qubits used~\cite{CN}. It is however sensible to minimize the number of qubits before applying error correction if qubits are hard to come by.

\section{Conclusion}
\noindent
Putting together several tricks, we have developed a circuit for the quantum part of the factorization algorithm, that is, the order-finding algorithm, while focusing on reducing the number of qubits. The number of qubits needed is $2n+3$ and the depth is $O(n^3)$. This circuit uses slightly less qubits than those previously known if restricted to elementary gates. It is also completely general and does not rely on any properties of the number to be factored.

Given the values $a$ and $N$, this circuit gives the order $r$ of $a$ modulo $N$ with good probability. Many runs of this algorithm may be needed to factor a number. Also, the randomly chosen value $a$ is hardwired in the circuit and there is a probability (about one half) that it will be necessary to choose a new value $a$ and run a new order-finding algorithm on it. This is not a problem if the quantum computer is a physical device where the gates are interactions controlled by a classical computer such as laser pulses on trapped ions, NMR and most implementation proposals. Indeed, the circuit can easily be classically computed. A quantum computer consisting of a physical system controlled by a classical computer is the most conceivable option at this point.

\section*{Acknowledgements}
\noindent
The author is very grateful to Michel Arsenault for many helpful comments and corrections, and would also like to thank Jos\'{e} Manuel Fernandez, Christof Zalka and Anne Broadbent. This work was supported in part by CSE of Canada, and the author acknowledges support from NSERC of Canada.


\end{document}